\documentclass{PoS}

\usepackage{wrapfig}
\usepackage{subfigure}
\usepackage{amsmath, amssymb, amsfonts}
\usepackage{eurosym}

\newcommand{\Tr}{\operatorname{Tr}}
\newcommand{\ee}{\ensuremath{\textrm{e}}}
\newcommand{\abs}[1]{\lvert {#1}\,\rvert} 

\title{LatticeQCD using OpenCL}

\ShortTitle{LatticeQCD using OpenCL}

\author{Owe Philipsen, \speaker{Christopher Pinke}, Christian Schäfer, Lars Zeidlewicz\\
        Institut für Theoretische Physik - Johann Wolfgang Goethe-Universität\\
        Max-von-Laue-Str. 1, 60438 Frankfurt am Main\\
        E-mail: \email{philipsen, pinke, cschaefer, zeidlewicz @th.physik.uni-frankfurt.de} }

\author{{Matthias Bach}\\
        Frankfurt Institute for Advanced Studies / Institut für Informatik - Johann Wolfgang Goethe-Universität\\
				Ruth-Moufang-Str. 1, 60438 Frankfurt am Main \\
        E-mail: \email{bach@compeng.uni-frankfurt.de}}

\abstract{We report on our implementation of LatticeQCD applications using OpenCL. We focus on the general concept and on distributing different parts on hybrid systems, consisting of both CPUs (Central Processing Units) and GPUs (Graphic Processing Units).}

\FullConference{XXIX International Symposium on Lattice Field Theory \\
                 July 10 - 16 2011\\
                 Squaw Valley, Lake Tahoe, California}

\begin{document}

\section{Introduction}

\begin{wrapfigure}{r}{0.4\textwidth}
 \centering
 \includegraphics[scale=.25]{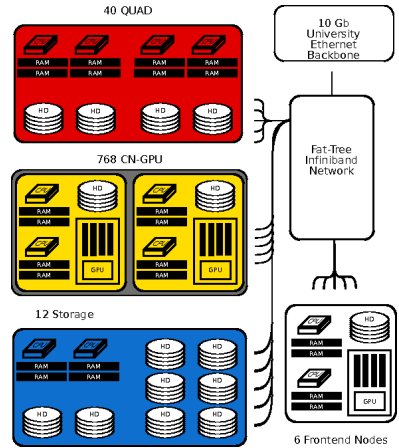}
 \caption{LOEWE-CSC}
 \label{loewe-arch}
\end{wrapfigure}
Graphic Processing Units (GPUs) offer a computing architecture well suited for LatticeQCD applications. Consequently, there is an on-going software- and algorithm development in order to incorporate GPUs effectively into lattice simulations. See for example \cite{Bonati:2011dv, Babich:2010mu, Clark:2009wm, Babich:2011np}. These applications are developed and carried out predominantly on NVIDIA hardware, consistently using the NVIDIA exclusive CUDA language \cite{cudaurl} for the interaction with the GPU. Using GPUs is attractive because they have good price-per-flop (e.g. $\approx$ 2,0 \geneuro / Gflop on a NVIDIA GTX 580 and even $\approx$ 0,4 \geneuro / Gflop on an AMD 6970 \cite{geizhals}) and performance-per-watt ratios.

Recently, a new cluster was introduced at Frankfurt University, the ``LOEWE-CSC'' \cite{loewecscurl}. It is dedicated to high-performance computing, but contrary to existing clusters it solely consists of AMD hardware. A sketch of its infrastructure is shown in Fig. \ref{loewe-arch}. A striking feature is its heterogeneous architecture: The majority of its compute nodes each hold two 12-core AMD Magny-Cours Central Processing Units (CPUs) and one AMD RADEON 5870 GPU. The LOEWE-CSC is ranked 22 in the Top500 list of supercomputers \cite{top500url} and rank 10 in the Green500 list of energy-efficient supercomputers (with 718 Mflops/Watt) \cite{greentop500url}.

However, presently existing GPU appplications are mostly suitable for NVIDIA hardware. Other than using graphic application programming interfaces (APIs) like OpenGL \cite{openglurl}, the only tool available to use AMD GPUs for general purposes is OpenCL \cite{openclurl}. This is an open standard for parallel programming. Furthermore, it is explicitly designed for heterogeneous (or hybrid) systems, thus being well suited for the LOEWE-CSC as well as other, non-GPU platforms. Implementations of OpenCL can be found both from AMD (AMD Accelerated Parallel Processing (APP) \cite{amdappurl}, formerly ATI Stream SDK) and NVIDIA (as part of CUDA). 

The first lattice simulations in OpenCL were performed in \cite{Bonati:2011dv} with staggered fermions. On NVIDIA hardware, a significantly lower performance (25\% on C1060 and 60\% on S2050) of OpenCL was reported compared to CUDA for Hybrid Monte Carlo (HMC) updates. An AMD GPU was also considered. Here, OpenCL performance is better than on Nvidia hardware, but still below CUDA results (50\% less performance on an AMD 5870 in OpenCL than on a S2050 in CUDA).

\section{OpenCL}
\label{opencl}

\begin{wrapfigure}{r}{0.35\textwidth}
\vspace{-5pt}
 \centering
 \includegraphics[scale=.26]{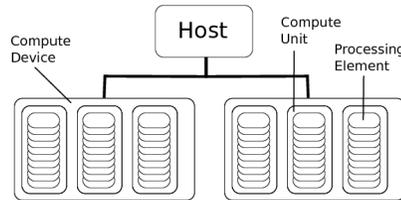}
 \caption{OpenCL Concept}
\vspace{-5pt}
 \label{opencl_concept}
\end{wrapfigure}

In the following, we will present the general ideas of OpenCL and explain important terms. For more information see \cite{openclstd}. The generic concept of an OpenCL application consists of a ``host`` program and several ``compute devices`` (see Fig. \ref{opencl_concept}). They live together on a so called ``platform``.  The host controls memory management and calculations carried out on the devices, while each device may either be a GPU, a CPU or any other kind of supported compute device. A single device then consists of a bunch of ''compute units`` (on a multicore CPU this would be a single core) which in turn can consist of one or more ''processing elements`` that carry
 out the actual computations. This of course reflects the architecture of GPUs. On CPUs a compute unit (i.e. a single core) and a processing element may also be the same, although this depends on the specific model and OpenCL implementation used. It should be noted that on a CPU it is possible to split up a multicore CPU into several devices, effectively grouping the cores suited for specific tasks.

Furthermore, execution commands for OpenCL functions (''kernels'') are scheduled in one or more ``command queues'' via the host. The queue launches execution of kernels on a specific device and also handles memory commands. Synchronization on this level is possible only within a command queue.

Central objects of any OpenCL application are the kernels. These have to be written in the \texttt{OpenCL C} programming language, which is based on a subset of \texttt{C99}, the \texttt{C} standard. Optionally, ``native kernels`` can be included from libraries. Kernels are executed using an up to 3-dimensional index space, where each index can be mapped on an instance of the kernel, which in turn are called ''work items''. Several work items make up a ''work group``, which allows for synchronization between work items. Kernels can access memory on various levels, ranging from ''global'' (e.g. the main memory on the GPU) to ``private'' (e.g. General Purpose Registers (GPRs) of a GPU stream-core). Besides the nomenclature, the setup is very similiar to CUDA. 

It is important to emphasize here that OpenCL allows for data- as well as for task parallel applications, meaning that it is on the one hand possible to perform SIMD computations and on the other hand to perform different (possibly independent) tasks in a parallel fashion, providing OpenMP \cite{ompurl} or UNIX's pthreads functionality automatically.

In order to have a hardware-independent programming model, the actual OpenCL program is compiled and built at runtime of the (host) application. This is done using an OpenCL inherent compiler. 

\section{Implementation}

The physical problems we are currently interested in are investigations of the quark gluon plasma (QGP) and the thermal transition of QCD with dynamical fermions \cite{Ilgenfritz:2009ns, Philipsen:2008gq, Burger:2011zc} as well as in pure gauge theory (PGT). These have yet been carried out mainly relying on the \texttt{tmlqcd} program suite \cite{Jansen:2009xp} and an application written with \texttt{QDP++} \cite{qdp++url}. Several features shall thus be provided by the OpenCL application: On the PGT side we need a SU(3) heatbath algorithm \cite{Creutz:1980zw}, whereas on the fermionic side we require an HMC algorithm including standard features (even-odd preconditioning, 2MN integrator, multiple integration timescales)  for $N_f = 2$ twisted-mass Wilson fermions \cite{Jansen:2009xp, Shindler:2007vp}. Also, ILDG-compatible I/O is required \cite{ildgurl}.

In order to account for the fundamentally different concept of OpenCL we decided to write an all new program. The implementation of the desired features mentioned above resulted in four executables, providing the possiblity to generate gauge configurations as well as calculate physical observables of interest for pure gauge theory and dynamical fermions. All calculations are performed in OpenCL. In the following we will go to some details describing the concrete implementation.

\begin{figure}[h]
	\begin{center}
		\subfigure[OpenCL module classes]{
			\centering 
 			\includegraphics[scale=0.7]{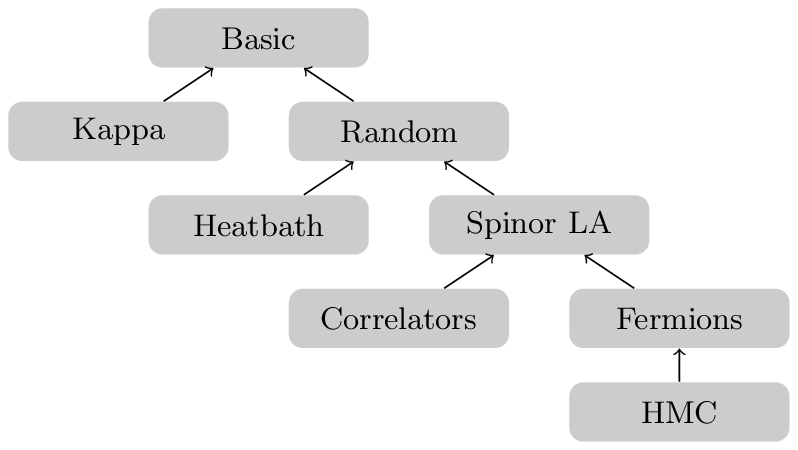}
		}
		\subfigure[Gaugefield classes]{
			\centering
			\includegraphics[scale=0.7]{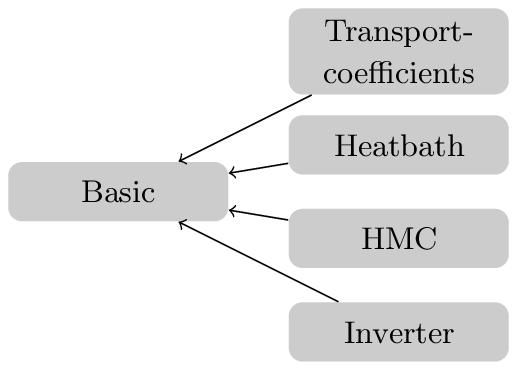}
		}
		\caption{Structure of opencl modules- and gaugefield classes}
		\label{prog_struct}
	\end{center}
\end{figure}

\vspace{-.8cm}

The host program was set up in \texttt{C++} in order to implement independent program parts easily using \texttt{C++} classes and to have extension capabilities in a natural way. 

The central object is the class \texttt{gaugefield}, which incorporates the initialization of OpenCL and holds an application-specific number of \texttt{opencl\_device}-objects. As the name indicates, the latter class contains all compute device-related parts, such as kernels or memory objects, and eventually executes the kernels on a specific device. The class \texttt{gaugefield} is also dedicated to synchronize the physical gauge field when it is used on several devices. 

The specific physical problems were implemented as child classes of \texttt{gaugefield} and \texttt{opencl\_module}, containing the problem-related functionality (see Fig. \ref{prog_struct}). Furthermore, for each physical problem a number of ``tasks'' can be defined which then again can contain a number of device objects to carry out this task. For example, the \texttt{inverter} executable essentially performs two tasks, the inversion of the fermion matrix and the calculation of correlators. This concept will prove useful when looking at hybrid applications.

\begin{wrapfigure}{r}{0.55\textwidth}
 \centering
	\vspace{-4mm}
  \includegraphics[scale=.6]{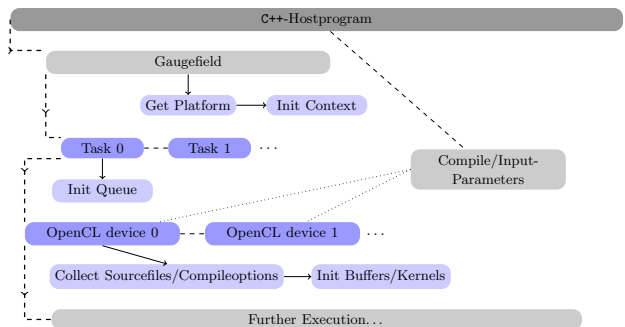}
 \caption{Schematic flow of program initialization}
 \label{prog-flow}
\end{wrapfigure}

As was mentioned in section \ref{opencl}, the OpenCL environment has to go through certain initialization processes before the actual calculations can be carried out. A schematic flow of this process is shown in Fig. \ref{prog-flow}. The major part of the initialization is spent on generating the kernels. This is done in a couple of steps: First, the files needed for the kernel code are collected, after that an OpenCL ``program`` is compiled and linked using the OpenCL compiler. This program can then be used to build kernels referring to functions declared with \texttt{\_\_kernel} within the previously read-in source code.

We found it convenient to build every kernel as a stand alone program since the kernel is the object of interest. The compiler generates binary files which can be used to extract informations about the kernel, e.g. GPR usage, which is useful for benchmarking and optimization. They can also be reused for kernel generation at a later program run. This speeds up the initialization time significantly.

One can influence the whole process at this point by means of the simulation parameters. The latter are typically read in at runtime of the host, so one can e.g. switch between CPU and GPU simply via the input file. Since the kernels are compiled only at runtime, one can pass the simulation parameters (e.g. \texttt{NT}, \texttt{NS}, \texttt{$\beta$}, \ldots) to them as compile time parameters. This is a nice way of avoiding many kernel arguments as well as to ``hard code'' the parameters into the kernel code.

\begin{figure}[h]
	\begin{center}
		\subfigure{
			\centering
			\includegraphics[scale=.57]{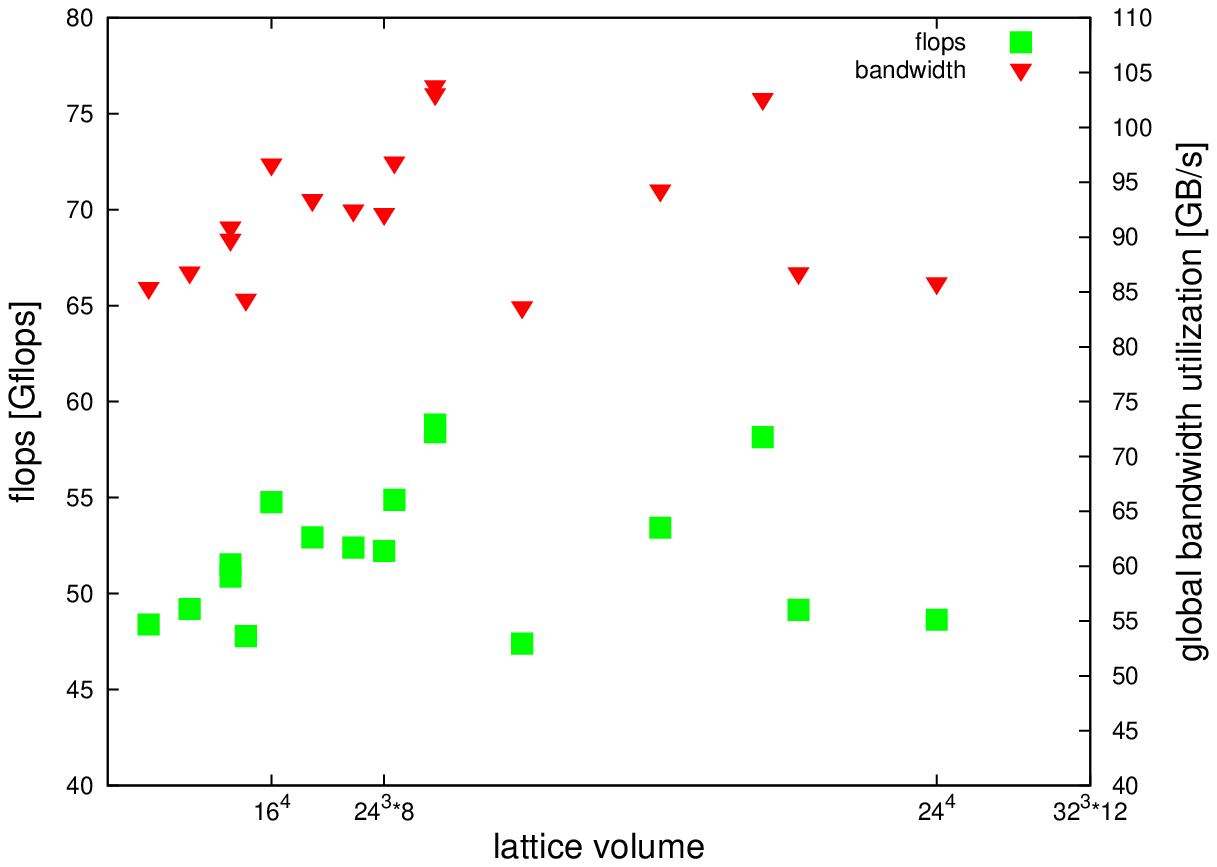}
		}
		\subfigure{
			\centering
			\includegraphics[scale=.57]{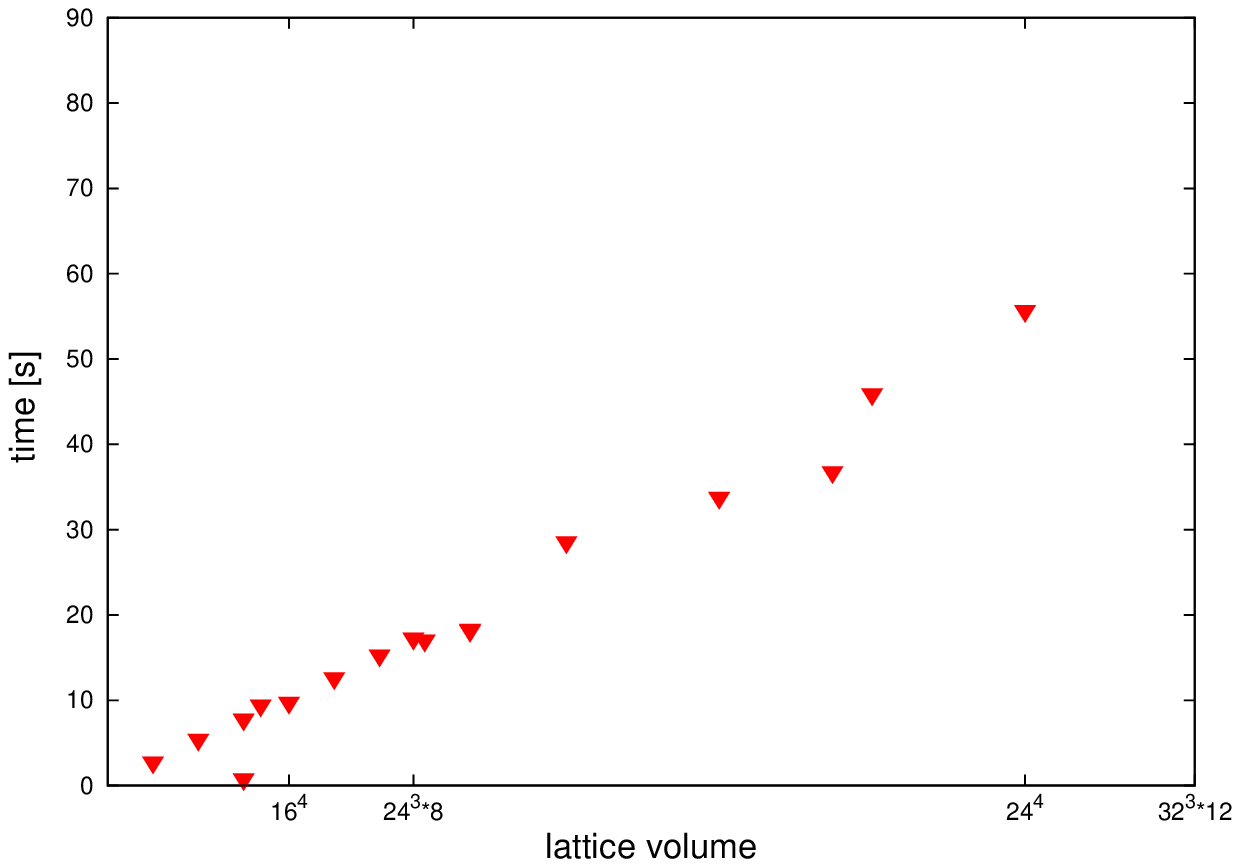}
		}
		 \caption{Dslash performance on AMD 5870 on various even-odd preconditioned lattices (statistics are of $\mathcal{O}(10k)$ for each volume).}
		\label{dslashperformance}
	\end{center}
\end{figure}

\vspace{-.8cm}

In fermionic applications, the most time-consuming part is the inversion of the fermion matrix and the non-diagonal part of the Dirac matrix (``dslash''), respectively. On GPUs, this problem is always bandwidth-limited and tuning is required to achieve a satisfiying amount of the maximum bandwidth (on an AMD 5870 this is e.g. 154 GB/s). Our (even-odd preconditioned) dslash implementation currently performs at 45 - 60 GFlops in double precision calculations (with a bandwidth utilization of up to 105 GB/s) over a wide range of lattice sizes (See Fig. \ref{dslashperformance}). We will give more performance results for AMD hardware (also considering memory optimizations like \texttt{RECONSTRUCT TWELVE} \cite{Clark:2009wm}) in a future publication.

\section{Hybrid strategies}

Having a hybrid system as the LOEWE-CSC at hand, the question arises how one can use this infrastructure effectively. Both GPU and CPU hold several advantages, qualifying them for certain tasks. GPUs outperform CPUs when it comes to floating point operations, whereas a CPU can in general operate a bigger amount of memory and a bigger cache, just to name a few. OpenCL can be used quite easily to distribute computations over a hybrid system.

\begin{wrapfigure}{r}{0.5\textwidth}
	\vspace{-5mm}
	\hspace{-15mm}
	\begin{center}
			\centering
			\includegraphics[scale=0.9]{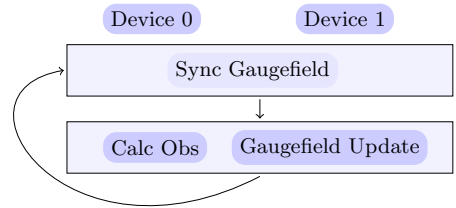}
		\caption{Simplest hybrid strategy.}
		\label{aaaaaaaaaaa}
	\end{center}
	\vspace{-5mm}
\end{wrapfigure}

A typical scenario in lattice simulations is the iterative calculation of some observables out of a sequence of gauge configurations.  The simplest case one can have is an observable that does not require any synchronization in between its calculation. Given 2 devices, one can distribute two generalized tasks among them, as depicted in Fig. \ref{aaaaaaaaaaa}. One task calculates the observable while in the meantime the other provides the ingredients required for the next iteration's calculation. Synchronization between the devices is carried out at the start of each iteration.

We implemented this concept for two observables, the $N_f = 2$ mesonic flavour doublet correlators with quantum number $\Gamma$, 
\begin{equation}
	C_\Gamma = - \Tr \left( S_u^\dagger(x_0,x) \gamma_5 \Gamma S_u (x_0,x) \Gamma \gamma_5 \right)\;,
\end{equation}
for which one has to provide the propagator $S_u (x_0,x) \sim M^{-1} b (x_0)$ (where $M$ is the fermion matrix and $b$ a point source at site $x_0$), and the second order transport coefficent $\kappa$ of the Quark Gluon Plasma \cite{Romatschke:2009ng}. $\kappa$ can be extracted from the retarted propagator $G_R$ at zero Matsubara frequency, 
\begin{equation}
G_R (\omega=0, \vec{q}) = G(0) - \frac{\kappa}{2}\abs{\vec{q}}^2 + \mathcal{O}(\abs{\vec{q}}^3) \;,
\end{equation}
which can be calculated on a given gauge configuration by the euclidean correlator $G_E$ according to
\begin{equation} 
	G_R (\omega=0, \vec{q}) =  G_E (\omega=0, \vec{q})  = N\; \sum_{x,y} \ee^ {q_3(x_3 - y_3)} \left< T_{12}(x)T_{12}(y) \right>\;.
\end{equation}
$T_{\mu\nu}$ is a discretization of the energy-momentum tensor using clover-plaquettes \cite{Maezawa:2010sq}.

\begin{figure}[h]
	\begin{center}
		\subfigure[Mesonic correlator $C_\Gamma$]{
			\centering
			\includegraphics[scale=0.9]{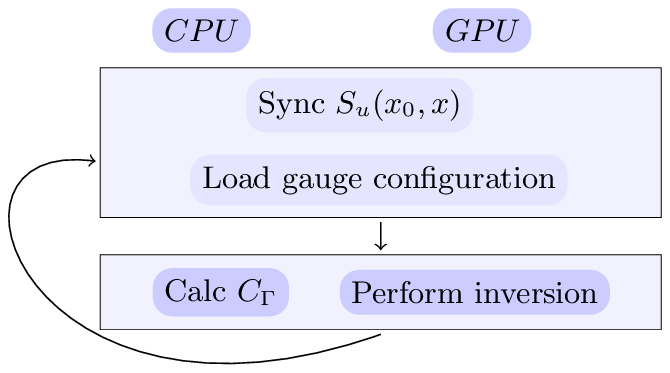}
		}
		\subfigure[Transportcoefficient $\kappa$]{
			\centering
			\includegraphics[scale=0.9]{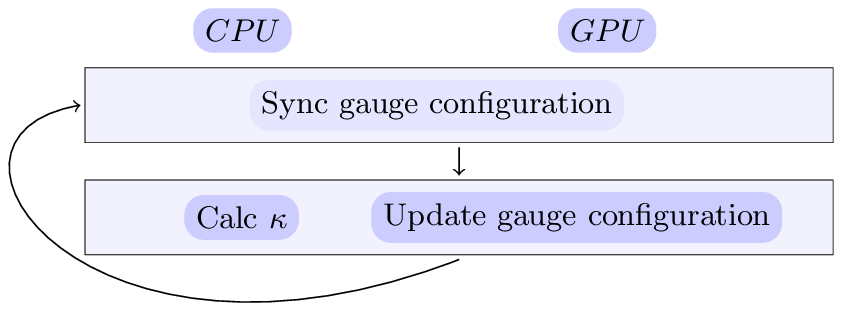}
		}
		\caption{Hybrid strategies for two observables of interest.}
		\label{hyb_str_obs}
	\end{center}
\end{figure}

The concrete implementations can be seen in Fig. \ref{hyb_str_obs}, where we considered the case of one CPU and one GPU device in the system (note that e.g. on one LOEWE-CSC node, one CPU device has by default 2*12 = 24 cores). The assignments of the different tasks to the devices came quite naturally, since $\kappa$'s calculation requires much more memory ressources than the heatbath algorithm and the inversion of the fermion matrix is carried out faster on the GPU. 

\section{Conclusions}

We presented to some detail our implementation of LatticeQCD applications using OpenCL. This is a quite different approach compared to other applications around, which mainly operate on NVIDIA hardware using CUDA. Since OpenCL is a hardware-independent programming model, any hardware can be used, which offer an optimal price-per-flop ratio.
Especially, parallel calculations can be performed quite simply in OpenCL, allowing for applications suited for hybrid architectures. We gave two examples using GPU and CPU devices at the same time effectively.
We are currently benchmarking and optimizing our code to exploit the compute powers of the LOEWE-CSC and AMD hardware in general, providing an alternative to NVIDIA hardware bound applications.

\acknowledgments

O. P. and C. P. are supported by the German BMBF grant \textit{FAIR theory: the QCD phase diagram at vanishing and finite baryon density}, 06MS9150. L.Z. is supported by the DFG grant \textit{phase transition and screening masses in $N_f = 2$ QCD}, PH 158/3-1. M. B., O. P, and C. S. are supported by the Helmholtz International Center for FAIR within the LOEWE program of the State of Hesse. M.B. is supported by the GSI Helmholtzzentrum für Schwerionenforschung. C.P. acknowledges travel support by the Helmholtz Graduate School HIRe for FAIR.


\bibliographystyle{/bibstyles/hieeetr.btr}

\end{document}